\documentclass[showpacs,9pt,aps,prd,reprint,onecolumn,notitlepage,superscriptaddress]{revtex4-1}

\usepackage{amsmath}
\usepackage{bbm}
\usepackage{amsfonts}
\usepackage{amssymb}
\usepackage{bm}
\usepackage{tensor}
\usepackage{graphicx}
\usepackage[]{subfigure}
\usepackage{color}

\usepackage[hidelinks]{hyperref}
\hypersetup{
    bookmarks=true,         
    unicode=true,          
    pdftoolbar=true,        
    pdfmenubar=true,        
    pdffitwindow=true,     
    pdfstartview={FitH},    
    pdftitle={My title},    
    pdfauthor={author},     
    pdfsubject={Subject},   
    pdfcreator={Creator},   
    pdfproducer={Producer}, 
    pdfkeywords={keyword1} {key2} {key3}, 
    pdfnewwindow=true,      
    colorlinks=true ,       
    linkcolor=blue,       
    citecolor=blue,      
    filecolor=green,      
    urlcolor=black           
}

\providecommand{\ed}{\mathrm{d}}
\providecommand{\Horava}{Ho\v{r}ava~}
\providecommand{\lu}{{l_{\uparrow}}}
\providecommand{\ld}{{l_{\downarrow}}}
\providecommand{\mlu}{{\mathbf{l}_{\uparrow}}}
\providecommand{\mld}{{\mathbf{l}_{\downarrow}}}
\providecommand{\tu}{\Theta_{\uparrow}}
\providecommand{\td}{\Theta_{\downarrow}}
\providecommand{\br}{\bar{r}}
\providecommand{\lieu}{\mathcal{L}_{\uparrow}}
\providecommand{\lied}{\mathcal{L}_{\downarrow}}
\providecommand{\ah}{\textrm{\tiny{AH}}}
\providecommand{\eh}{\textrm{\tiny{EH}}}
\providecommand{\arctanh}{\textrm{arctanh}}

\begin{document}

\title{Evolving \Horava Cosmological Horizons}

\author{Mohsen Fathi}
\email{m.fathi@shargh.tpnu.ac.ir;\,\, mohsen.fathi@gmail.com}

\author{Morteza Mohseni}
\email{m-mohseni@pnu.ac.ir}

\affiliation{Department of Physics, Payame Noor University (PNU), P.O. Box 19395-3697 Tehran, I.R. of IRAN
}

\begin{abstract}
Several sets of radially propagating null congruence generators are exploited in order to form 3-dimensional marginally trapped surfaces, referred to as black hole and cosmological apparent horizons in a \Horava universe. Based on this method, we deal with the characteristics of the 2-dimensional space-like spheres of symmetry and the peculiarities of having trapping horizons. Moreover, we apply this method in standard expanding and contracting FLRW cosmological models of a \Horava universe to investigate the conditions under which the extra parameters of the theory may lead to trapped/anti-trapped surfaces both in the future and in the past. We also include the cases of negative time, referred to as the finite past, and discuss the formation of anti-trapped surfaces inside the cosmological apparent horizons.
\\\\
{\textit{keywords}}: \Horava universe, Standard cosmology, Null congruence expansion, Black hole horizons, White holes.

\end{abstract}

\maketitle

\section{Introduction}

When it comes to those ``small'' theories of quantum gravity which emerge from super strings, the 4-dimensional \Horava theory, proposing a quantum field model including the UV completion of the Einstein theory and its relevance to anisotropies between space and time, becomes of great significance. {{Indeed, \Horava gravity deals with the differences between the concepts of space and time which rise from quantum theory, and makes these anisotropies appear at high energy levels.}} The \Horava gravity which was introduced in Ref.~\cite{Horava2009} includes the theoretical aspects of critical phenomena, analogous to those in  condensed matter physics. This theory leads back to Einstein's general relativity in infrared limits. {{However, the original formulation has had some shortcomings, like predicting extremely different results for slightly perturbed spherical geometries from those of complete spherical symmetry.}} Ever since it was revealed to the public in 2009, the theory has received great attention and has been followed by a tremendous number of works by other scientists. {{It has been shown that the theory suffers from some problems and inconsistencies. For example it exhibits strong couplings, explaining why solutions to general relativity are not recovered in this model \cite{Charmousis2009}.\\

One important version of \Horava gravity (which we consider in this paper) is the projectable one, where the lapse is assumed to be only a function of time. This version has also some problems, like the existence of an unstable scalar degree of freedom, which once again leads to strong couplings. However it has been shown that this version can be regarded as a useful tool in developing the theory, because it greatly reduces the number of terms in the action \cite{Weinfurtner2010} and therefore, illuminates the way of looking deeply on the variations of \Horava gravity. It is worth mentioning the arguments raised by some scientists, that projectable versions may become useless in infrared limits, because in those limits quantum corrections become controllable \cite{Koyama2010}. In terms of consistency, the covariance breaking of the original \Horava model has been examined to argue that the resultant scalar degree of freedom may constitute an extra mode which has been proved to perform as the strong coupling \cite{Blas2009}. \\

To overcome these problems, an extended model of \Horava gravity was proposed in Ref.~\cite{Blas2010}, where using a regular quadratic action, the renormalizability of the scalar mode is recovered. Moreover, studying the Hamiltonian formulation of $f(R)-$\Horava theories, Kluso\v{n} proposed them as some ``healthy'' extensions of \Horava gravity \cite{Kluson2010a, Kluson2010b}. Furthermore, it  has been shown that by imposing a local $\mathrm{U}(1)$ gauge symmetry, one can eliminate the mentioned scalar mode and retain the general covariance and hence, another healthy version of the theory \cite{Horava2010}. This method was also applied to the $f(R)-$\Horava theories to obtain their covariant extension \cite{Kluson2011a}. The quest for healthy variations is continued. For example, it has been proved that the degrees of freedom will reduce to those of general relativity, if a Lagrange multiplier is imposed on the \Horava action \cite{Kluson2011b}.\\

In order to alter the projectability condition, the holographic implications of \Horava gravity and its duality to quantum field theory with anisotropic Lifshitz scaling have been discussed in Ref.~\cite{Griffin2013}. There, the authors introduce some classes of quantum field theories, based on the ``unhealthy reduction" of non-projectable \Horava gravity. Furthermore and in Ref.~\cite{Chaichian2015}, such non-projectable classes were shown to be capable of becoming free of unwanted extra scalar modes. This therefore provides us a healthy extension of the non-projectable \Horava gravity, whose renormalizability has been addressed in Ref.~\cite{Barvinsky2016}.\\

Despite the fact that the abilities and weaknesses of \Horava gravity have been under debate, the original projectable theory has proved to be of great interest, especially when it was endowed with some black hole solutions. This therefore led to rigorous discussions on the concept of \Horava black holes.}} These solutions by Park \cite{Park2009}, Kehagias and Sfetsos \cite{Kehagias2009}, and Lu, Mei and Pope \cite{Lu2009}, and for example the solution classes introduced in Ref.~\cite{Capasso2010}, constitute the black hole spacetimes and makes it possible to go further in studying the horizon thermodynamics ({{actually the dynamics of apparent horizons in a Friedmann-Lama\^{i}tre-Robertson-Walker (FLRW) universe and its relation to thermodynamics has been investigated in Refs.~\cite{Cai2005, Cai2009c} and for example the relevant thermodynamics of the topological black-holes introduced in Ref.~\cite{Cai2009a}, have been discussed in Refs.~\cite{Cai2009b, Cai2010}}}) and the energy-momentum conditions of such black holes (Refs.~\cite{Radinschi2011, Radinschi2012}). Such static spacetimes have even been considered in characterizing the behaviour of time-like objects while orbiting a \Horava black hole (e.g. Refs.~\cite{Chen2010, Abdujabbarov2011, Sadeghi2012}).\\

However, when we are asked about the most significant feature of black holes, we definitely bring up the concept of singularities and the cosmic censorship \cite{Penrose1965, Hawking1970}. This means that there will be regions beyond the ``hole's" event horizon which are causally disconnected from any observer outside the event horizon. Therefore what is of crucial importance in studying a hole is  its horizons. Mathematically, in a 4-dimensional spacetime, an event horizon is a 3-dimensional closure of a surface, on which a congruence of ingoing and outgoing null flows will converge. In this regard, the event horizon is itself generated by the null congruences. On the other hand, the full perception of event horizons relies on complete knowledge about future null infinity, which is impossible to  achieve. Hence, to talk about the physical phenomena of holes, one should talk about other types of horizons that the hole may possess; such as apparent horizons.  Indeed, apparent horizons are those geometric features of holes on which their thermodynamics is studied. Apparent horizons are 3-dimensional hypersurfaces which are foliated by marginally trapped surfaces for null congruence. Therefore, since apparent horizons are not necessarily separating two causally disconnected regions, they are more accessible and are of great proficiency in gravitational treatments of holes. Such horizons may even evolve for a cosmological hole which is indeed a hole on a cosmological background. In that case, the spacetime splits up itself into hole horizons and cosmological horizons. Therefore it is important to note that a cosmological horizon can evolve in time. In this regard, time-dependent hole horizons are analogous to an expanding or contracting FLRW cosmological horizon, whereas time-independent static Schwarzschild horizons are analogous to de Sitter static cosmological horizons. Hence, the concept of evolving cosmological and hole horizons is an interesting topic of research in studying possible black holes which evolve on a cosmological background, in every peculiar theory of gravitation and their corresponding cosmological models. \\

In the case of a \Horava universe, which lies within the scope of this paper, there have been some efforts to build cosmological black/white holes upon scalar field constituents of the theory and time-like flows of the cosmic fluid. For example in Ref.~\cite{Barausse2011}, the authors compare the black hole solutions in the infrared limit of \Horava gravity which has been defined on a time-like \ae ther, to those known Schwarzschild-like black holes. Also in Ref.~\cite{Afshordi2014a}, it has been shown that the usual McVittie cosmological black hole is also a \Horava black hole, in the special case of the Lorentz violating parameter to be 1/3. The \Horava theory has also proved to generate universal horizons, emerged from scalar couplings in the theory (see Ref.~\cite{Afshordi2014b}). \\

In this paper, we investigate the possibilities of the formation of trapped/anti-trapped surfaces inside cosmological apparent horizons, and their corresponding conditions imposed on the specific parameters which appear in the equations of motion and spacetime geometries. In particular, we justify the following inferences:

\begin{itemize}

\item{Null congruence generators on static \Horava black hole spacetimes confirm that each of them possess only one event horizon, separating time-like and space-like regions; they correspond only to the formation of one type of trapping horizon in the future.}

\item{On a FLRW cosmological background, we realize both time-like and space-like apparent horizons residing respectively inside and outside event horizons. {{The existence of space-like apparent horizons in FLRW spacetime is claimed to be a consequence of the existence of exotic matter. Actually, in de Sitter spacetime (vacuum universe with constant Hubble parameter) one can expect space-like apparent horizons. However we show that for a \Horava universe this is the case even for a variable Hubble parameter. Indeed in a vacuum \Horava universe on a FLRW background, space-like apparent horizons can form and evolve.}} }

\item{These apparent horizons evolve in time, but make a remarkable difference in their respectable relevance between the field equation constants and the spatial curvature of space.}

\item{Expanding and contracting \Horava universes will actually distinguish between the types of trapping horizon. Moreover, their relations to the evolution in the finite past indicates formations of trapped/anti-trapped surfaces within negative time. }

\end{itemize}

The paper is organized as follows: In Section~\ref{sec:Horava}, we briefly highlight important traits of \Horava gravity and its proposed black hole solutions. In Section~\ref{sec:nullstatic}, we introduce our method of treating null congruence expansion on Kehagias-Sfetsos (KS) and Lu-Mei-Pope (LMP) static black holes and characterize the space-like surfaces to foliate 3-dimensional trapping horizons. In Section~\ref{sec:evolving} the LMP cosmological solution in a vacuum \Horava universe is employed in order to inspect the evolution of time-like and space-like apparent horizons. We discuss the categories according to which these temporal evolutions are characterized. We conclude in Section~\ref{sec:conclusion}. Note that we choose the geometric units such that $c=G=1$.

\section{\Horava Theory and Its Static Black Hole Spacetimes}\label{sec:Horava}

Here we recap some important basics of \Horava gravity. A very good review can be found in Ref.~\cite{Mukohyama2010}. \Horava gravity is characterized by the 4-dimensional Arnowitt-Deser-Misner (ADM) formulation of general relativity and therefore by the 4-dimensional metric \cite{ADM1962}
\begin{equation}\label{eq:metric-general}
\ed s^2=-N(t)^2\ed t^2+g_{ij}(t,\vec{x})\left(\ed x^i+N^i(t,\vec{x})~\ed t\right)\left(\ed x^j+N^j(t,\vec{x})~\ed t\right),
\end{equation}
with $N(t)$, $N^i(t,\vec{x})$ and $g_{ij}(t,\vec{x})$ being the lapse, shift and  3-dimensional induced metric respectively. The lapse depends solely on the time, because of the projectability condition. The \Horava theory introduces a class of 3+1-dimensional modifications to general relativity. Accordingly, the UV completion action of the Einstein-Hilbert action would be
\begin{equation}\label{eq:UVcompletion}
I=\frac{1}{\kappa^2}\int N(t)~\ed t~\int\sqrt{g(t,\vec{x})}~\left(K^{ij}(t,\vec{x})K_{ij}(t,\vec{x})-\lambda K(t,\vec{x})^2\right)\ed^3\vec{x},
\end{equation}
in which the extrinsic curvature $K_{ij}$ is defined in terms of differentiations of the 3-dimensional metric with respect to the time coordinate,
\begin{equation}\label{eq:extrinsic}
K_{ij}(t,\vec{x})=\frac{1}{2N(t)}\left(\dot g_{ij}(t,\vec{x})-\nabla_i N_j(t,\vec{x})-\nabla_j N_i(t,\vec{x})\right).
\end{equation}
The Lorentz violating parameter $\lambda$ is a constant, but can have alternative values in different approaches. Note that in general relativity $\lambda=1$. Furthermore in Eq.~(\ref{eq:UVcompletion}), $K=K\indices{^i_i}$ and the dot in Eq.~(\ref{eq:extrinsic}) stands for differentiation with respect to $t$. The extrinsic curvature constitutes the kinetic term of the action. One remarkable issue is that $K_{ij}$ is covariant under diffeomorphisms that include gauge symmetries to preserve foliations. Accordingly, the coupling constant $\lambda$ means that both terms in the parenthesis in Eq.~(\ref{eq:UVcompletion}) are separately invariant under the mentioned diffeomorphisms. Based on this coupling constant, the generalized De Witt metric can be written as \cite{Horava2009}
\begin{equation}\label{eq:DeWitt}
G^{ijkl}=\frac{1}{2}\left(g^{ik}g^{jl}+g^{il}g^{jk}\right)-\lambda~g^{ij}g^{kl},
\end{equation}
which can be used to rewrite the 3+1-dimensional UV completion action as
\begin{equation}\label{eq:UVcompletion-new}
I_{\textrm{UV}}=\frac{\kappa^2}{8\kappa_w^4}\int N~\ed t~\sqrt{g}\left(R^{ij}-\frac{1}{2}Rg^{ij}+\Lambda_w g^{ij}\right)~\mathcal{G}_{ijkl}\left(R^{kl}-\frac{1}{2}Rg^{kl}+\Lambda_w g^{kl}\right)~\ed^3\vec{x},
\end{equation}
where the subscript $w$ refers to some 3-dimensional action in which the theory is ``detailed balanced". In the case of having Euclidean isotropic symmetry, the action $w$ is of the form of Einstein-Hilbert action
\begin{equation}\label{eq:E-L}
w=\frac{1}{\kappa^2_w}\int\sqrt{g}\left(R-2\Lambda_w\right)~\ed^3\vec{x}.
\end{equation}
Note that in Eq.~(\ref{eq:UVcompletion-new}), $R^{ij}$ is the 3-dimensional Ricci tensor $(R=R\indices{^i_i})$, and $\mathcal{G}_{ijkl}$ is the inverse of the De Witt metric in Eq.~(\ref{eq:DeWitt}).\\

Now for the case of $\lambda\neq1/3$, one can consider a static spherically symmetric solution to the potential in Eq.~(\ref{eq:UVcompletion-new}) of the form
\begin{equation}\label{eq:generalMetric}
\ed s^2=f(r)~\ed t^2+f(r)^{-1}\ed r^2+r^2\ed \Omega_{(2)}^2,
\end{equation}
where $\Omega_{(2)}$ constitutes the 2-sphere of symmetry. Based on this, the KS asymptotically flat black hole solution for $\lambda=1$ and $\Lambda_w=0$ is \cite{Kehagias2009}
\begin{equation}\label{eq:KS}
f(r)=1+\omega r^2-\sqrt{r\left(\omega^2 r^3+4\omega M\right)},
\end{equation}
with $M$ being a constant, $\omega=16\mu^2/\kappa^2$, and $\mu$ is the mass of dimensions $[\mu]=1$. Moreover, including the cosmological constant $\Lambda_w=2\Lambda/3$, then for the case of $\lambda>1/3$ and together with Eq.~(\ref{eq:generalMetric}), the LMP solution is obtained to be \cite{Lu2009}
\begin{equation}\label{eq:LMP}
f(r)=1-\Lambda_w r^2-\frac{\alpha}{\sqrt{-\Lambda_w}}\sqrt{r}.
\end{equation}
This solution requires $\Lambda_w<0$, however as we will see in the following sections, the cosmological implementations of the solution can be made for both cases of positive and negative $\Lambda_w$. Note that the significance of the case of $\lambda=1/3$ and the Weyl symmetry of the resultant theory have been highlighted by \Horava himself. \\

Now that we have mentioned the static black hole spacetimes defined in a \Horava universe, in the next section we use the method of constructing trapped horizons by means of null congruence and discuss the relevant conditions. From now on, all indices are 4-dimensional.

\section{Null Congruence Expansion on Static \Horava Black Holes}\label{sec:nullstatic}

In this section, we take KS and LMP black hole spacetimes to retrieve the black hole apparent horizons and the space-like 2-surfaces to foliate those horizons. To do this we take two null congruences corresponding to outgoing and ingoing trajectories. Letting $g_{ab}$ be the background spacetime metric, we can then generate these congruences by the tangential vector fields $\mlu$ and $\mld$, which constitute a transverse 2-surface described by the 2-metric \cite{Faraoni2013}
\begin{equation}\label{eq:2-metric}
h_{ab}=g_{ab}-\frac{\lu_a\ld_b+\lu_b\ld_a}{\lu^c\ld_c},
\end{equation}
in such a way that $h_{ab}\lu^b=\textbf{0}$; the congruence is orthogonal to the 2-surface described by $h_{ab}$. Also since the congruences are supposed to be null, $\lu^a\lu_a=\ld^a\ld_a=0$. Accordingly, the congruence expansion for outgoing and ingoing congruence on the so-called transverse surface would be \cite{Faraoni2013}
\begin{equation}\label{eq:expansion-general}
\Theta_{\updownarrow}=h^{ab}\nabla_a l_{\updownarrow b}.
\end{equation}
Note that $h_{ab}$ can also serve as the transverse projector, to project every kinematical characteristic of the congruence onto the transverse 2-surface (for full details see Ref.~\cite{Poisson2004}). In the following and by examining KS and LMP \Horava black holes, we show how the transverse expansion in Eq.~(\ref{eq:expansion-general}) is related to apparent and trapping horizons and the cosmic censorship of a black hole singularity.

\subsection{KS Static Black Holes}\label{sebsec:KS}

\begin{figure}[h]
\center{\includegraphics[width=8cm]{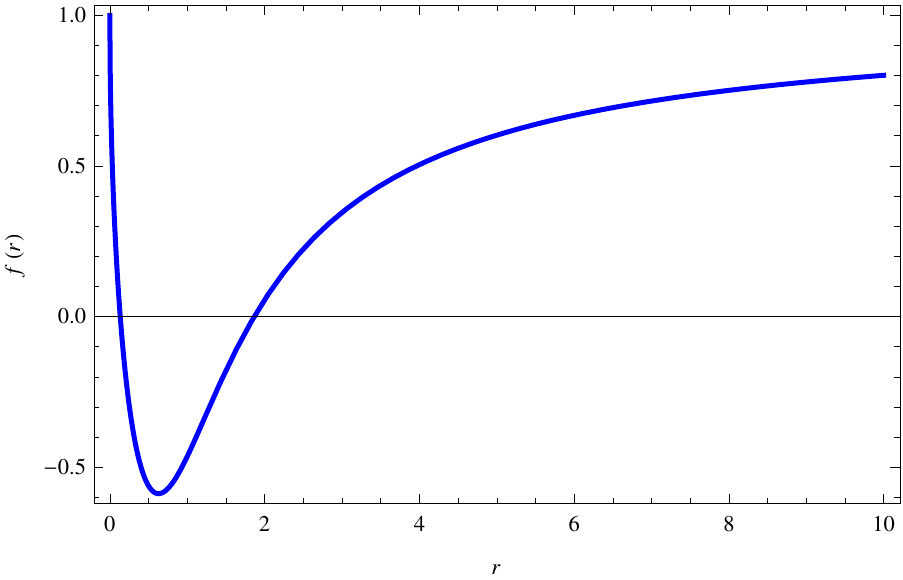}
\caption{\label{fig:f(r)-KS}The radial behaviour of KS metric potential, for $\omega=2$. The unit value along the $r$ axis is $M$.}}
\end{figure}

The KS black hole spacetime in Eq.~(\ref{eq:KS}) becomes irregular for $f(r)=0$ and the radii
\begin{equation}\label{eq:rpm-KS}
r_\pm=M\left(1\pm\sqrt{1-\frac{1}{2\omega M^2}}~\right),
\end{equation}
for which, in order to avoid a naked singularity, we must have $2\omega M^2\geq1$. Note that the true singularity is at $r=0$, where the scalar curvature diverges. Now to construct an  outgoing radially propagating null congruence generator, we write the light cone structure straight from the general spherical metric in Eq.~(\ref{eq:generalMetric}) as
\begin{equation}\label{eq:lightcone-spherical}
\epsilon=-f(r)~(\lu^t)^2+f(r)^{-1}(\lu^r)^2,
\end{equation} 
from which, according to the null condition $\epsilon=0$, we get
\begin{equation}\label{eq:outgoing-general}
\lu^a=\left(1,f(r),0,0\right).
\end{equation}
Also for $\mlu$ and $\mld$ to be normalized to -2, the ingoing null congruence generator becomes
\begin{equation}\label{eq:ingoing-general}
\ld^a=\left(\frac{1}{f(r)},-1,0,0\right).
\end{equation}
The outgoing and ingoing congruences expand according to Eq.~(\ref{eq:expansion-general}), giving
\begin{subequations}\label{eq:expansion-ingoing-outgoing-general} 
\begin{align}
\tu&=\frac{2f}{r},\label{eq:expansion-outgoing-general}
\\
\td&=-\frac{2}{r}.\label{eq:expansion-ingoing-general}
\end{align}
\end{subequations}
The black hole apparent horizons are foliated by marginally trapped 2-surfaces of outgoing congruence, on which $\tu=0$ and $\td<0$, of which the latter is automatically satisfied. Accordingly both radii $r_\pm$ give the location of apparent horizons. Therefore the configuration of the apparent horizons depends strictly on the choice of foliation. Indeed the apparent horizons constitute the closure of the 3-dimensional hypersurfaces
\begin{equation}\label{eq:hypersurface-apparent-general}
\Phi_\pm(r)=r-r_\pm=0.
\end{equation}
Note also that these apparent horizons are future horizons since $\td<0$ implies convergence of ingoing congruence in the future (the observer is outside the horizon). The normal vectors to the horizon hypersurfaces are tools which can  determine their characteristics. To work this out for the apparent horizons at $r_\pm$, let us define $\mathbf{N}_\pm$ to be the normal vectors to $\Phi_\pm$. We have
\begin{equation}\label{eq:AHNormal-general}
{N_\pm}_a=\partial_a\Phi_\pm=(0,1,0,0).
\end{equation}
Using Eq.~(\ref{eq:generalMetric}), it is straightforward to infer ${N_+}^a{N_+}_a={N_-}^a{N_-}_a=f(r)$. Peculiar to the case of KS black holes, the metric potential $f(r)$ has been plotted in Fig.~\ref{fig:f(r)-KS}. First of all, since $f(r_\pm)=0$, it is obvious that $\mathbf{N}_\pm|_{r=r_\pm}$ is null and hence, both KS apparent horizons are null hypersurfaces. On the other hand, $f(r)|_{r<r_-}>0$ and $f(r)|_{r>r_+}>0$, whereas $f(r)|_{r_-<r<r_+}<0$. This means that the normal vectors are space-like inside $r_-$ and outside $r_+$; those regions are time-like. Furthermore, since the normal vectors are time-like in the region between $r_-$ and $r_+$, that region is space-like. We can use these notions to argue that only one of the apparent horizons is also an event horizon. Essentially, since event horizons separate two causally disconnected regions, an event horizon must be null. As we showed above, this condition is automatically satisfied for $r_{\ah}=r_\pm$. Moreover, an event horizon must separate a time-like region from a space-like one. According to what we have mentioned above, $r_+$ is an event horizon and every congruence entering it has to converge on the inner horizon. However inside $r_-$ we encounter a time-like region and the congruence can avoid falling on the singularity, therefore $r_-$ cannot be regarded as an event horizon (this is clearly in contrast with what is stated by the authors of Ref.~\cite{Kehagias2009}). There is another way of looking at this by means of congruence expansion. According to Eq.~(\ref{eq:expansion-outgoing-general}), $\tu$ is negative in $r_-<r<r_+$, while it is positive in $r<r_-$ and $r>r_+$. This means that for the latter cases we have $\tu\td<0$, indicating un-trapped 2-surfaces. These foliating 2-surfaces of the apparent horizons become trapped between them, where $\tu<0$ and $\td<0$; the black hole region. Note that if we adopted a Killing vector $K^a=(1,0,0,0)$ of the spherically symmetric spacetime, then we could have pursued the same argument, based on the condition $K_a K^a=-f(r)$. Accordingly, both apparent horizons are also Killing horizons where $K^aK_a=0$. In this case, where $\mathbf{K}$ is time-like (outside $r_+$ and inside $r_-$) we encounter time-like regions and when it is space-like (inside the black hole region, $r_-<r<r_+$) we end up with a space-like region. 

\subsubsection{Trapping Horizons}\label{subsubsec:Trapping}

Trapped surfaces are the most indigenous traits of black holes. These surfaces are encompassed by 3-dimensional hypersurfaces known as trapping horizons. Now for a black hole apparent horizon to be also a trapping horizon, we should also take into account the condition $\lied\tu<0$, where $\lied$ denotes the Lie derivative in the direction of ingoing congruence. A hypersurface with such a condition is a future marginally outer trapped horizon (FMOTH) \cite{Faraoni2013}. Note that such a horizon becomes inner if $\lied\tu>0$. For the general spacetime in Eq.~(\ref{eq:generalMetric}) and the expansions in Eq.~(\ref{eq:expansion-ingoing-outgoing-general}), we have
\begin{equation}\label{eq:Lie-ingoing-general}
\lied\tu=\frac{2}{r^2}\left(f-rf'\right),
\end{equation}
which for the special case of KS black holes and according to Fig.~\ref{fig:f(r)-KS}, is negative on $r_+$ and positive on $r_-$. This means that the event horizon is a FMOTH whereas the inner apparent horizon is a FMITH. Inside FMOTH (or the event horizon) we expect to encounter trapped surfaces, as we discussed above.

\subsection{LMP Static Black Holes}\label{sebsec:LMP}

\begin{figure}[h]
\center{\includegraphics[width=8cm]{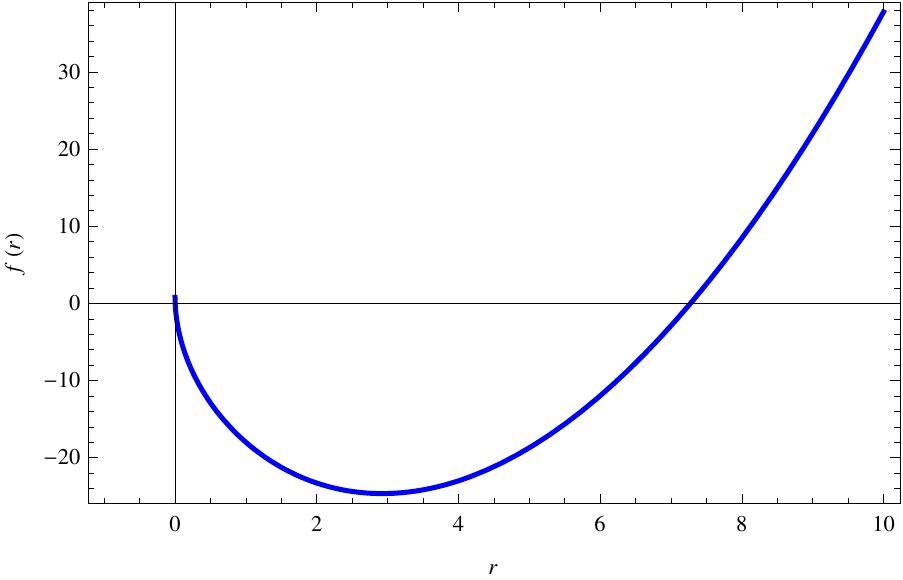}
\caption{\label{fig:f(r)-LMP}The radial behaviour of LMP metric potential for $\alpha=20$. The unit value along the $r$ axis is $-\Lambda_w$.}}
\end{figure}

The true singularity of LMP black holes defined in Eq.~(\ref{eq:LMP}) occurs at 
\begin{equation}\label{eq:LMP-trueSingularity}
r_{\tiny\textrm{sing}}=\frac{1}{4}\left(\frac{25\alpha^2}{-4\Lambda_w^3}\right)^{\frac{1}{3}},
\end{equation}
implying $\Lambda_w<0$. The radial behaviour of the LMP metric potential has been plotted in Fig.~\ref{fig:f(r)-LMP}.

According to the figure, $f(r)$ has only one non-negative solution $r_0$, indicating the irregularity of the spacetime. Hence Eq.~(\ref{eq:expansion-outgoing-general}) implies the possession of only one apparent horizon for LMP \Horava black holes. On the other hand, from Eq.~(\ref{eq:Lie-ingoing-general}) we infer that $\lied\tu<0$ holds in the whole case, and therefore $\Phi(r)=r-r_0=0$ is constructing a FMOTH. Moreover since $N_a=(0,1,0,0)$, the normal to the horizon is time-like outside $r_0$ and space-like inside it. Therefore $r=r_0$ is also an event horizon. This is in accordance with $\tu\td<0$ for $r>r_0$ (un-trapped surface), and $\tu<0$ and $\td<0$ for $r<r_0$ (trapped surface). The above notes show that in static spherically symmetric spacetimes, regardless of their number, all apparent horizons are fixed and if available, only one of them behaves as an event horizon.\\

Now that we have established our method of treatment, it is time to look for situations where we can deal with those horizons that can evolve in time. Such conditions are accessible within cosmological solutions of the gravitational theories. In what follows, we consider the LMP cosmological solutions of the \Horava universe and discuss the evolving apparent and trapping horizons.

\section{Evolving Horizons in a \Horava Universe}\label{sec:evolving}

As we saw in the previous section, when we consider black hole spacetimes, the ingoing congruence experiences a permanent convergence on the horizons in such a way that it never encounters anti-trapped surfaces inside the ``hole" region. One reason for this is that the gravitational structure is observed from outside the horizons. In other words, the observer will never detect outgoing congruence with permanent divergence; this property characterizes white hole regions and anti-trapped surfaces. So to provide such conditions, one way is to deal with observers inside the gravitational structure. This becomes possible when we consider an evolving (or fixed) cosmological system into which no ingoing congruence can enter. To elaborate this in a \Horava universe, in this section, we take the LMP cosmological solution for a vacuum universe and based on the method we introduced in Section~\ref{sec:nullstatic}, we look for the formation of anti-trapped surfaces encompassed by evolving apparent horizons.\\

For a FLRW perfect fluid cosmology described by the comoving line element
\begin{equation}\label{eq:FLRW}
\ed s^2=-\ed t^2+a(t)^2\left(\frac{\ed r^2}{1-k r^2}+r^2\ed\Omega_{(2)}^2\right),
\end{equation}
with $a(t)$ and $k$ the cosmic scale factor and the spatial curvature of the spacetime respectively, the LMP cosmological solutions for a vacuum \Horava universe have been calculated in Ref.~\cite{Lu2009} and are
\begin{subequations}\label{eq:LMP-cosmological} 
\begin{align}
H^2&=\frac{2}{3\lambda-1}\left(\frac{\Lambda_w}{2}-\frac{k}{a^2}+\frac{k^2}{2\Lambda_w a^4}\right),\label{eq:LMP-cosmological-H}
\\
\frac{\ddot a}{a}&=\frac{2}{3\lambda-1}\left(\frac{\Lambda_w}{2}-\frac{k^2}{2\Lambda_w a^4}\right),\label{eq:LMP-cosmological-q}
\end{align}
\end{subequations}
where $H=\dot a/a$. Accordingly, the explicit dependence of the scale factor on the cosmic time have been determined to be
\begin{equation}\label{eq:LMP-a(t)}
a(t)=\left(\frac{k}{\Lambda_w}+\alpha \textrm{e}^{2\sqrt{\frac{\Lambda_w}{3\lambda-1}}~t}~\right)^\frac{1}{2}.
\end{equation}
The significance of the solution in Eq.~(\ref{eq:LMP-a(t)}) is that it can be employed for either $\Lambda_w>0$, which is consistent with $\lambda>1/3$, or $\Lambda_w<0$ with $\lambda<1/3$.

\subsection{The Event Horizon}\label{subsec:EventHorizon}

The same method of constructing the light cone structure as we used in Eq.~(\ref{eq:lightcone-spherical}), can be applied to a universe described by the line element (\ref{eq:FLRW}). This way, we can obtain two radial null vectors
\begin{subequations}\label{eq:null-cosmic} 
\begin{align}
\lu^b&=\left(1,\frac{\sqrt{1-kr^2}}{a(t)},0,0\right),\label{eq:Lnull-cosmic-outgoing}
\\
\ld^b&=\left(1,-\frac{\sqrt{1-kr^2}}{a(t)},0,0\right),\label{eq:Lnull-cosmic-ingoing}
\end{align}
\end{subequations}
as generators of outgoing and ingoing congruences, which are normalized to $-2$. According to the ingoing congruence in Eq.~(\ref{eq:Lnull-cosmic-ingoing}), we have $\ld^1/\ld^0=dr/dt=-\sqrt{1-kr^2}/a(t)$. So if we assume that a congruence of ingoing rays which starts propagating at the Big Bang time $t_B$ and comoving radial position $r$ is being detected at time $t$ by an observer located at $r=0$, then we have $\int_r^0 \ed r'/\sqrt{1-kr'^2}=-\int_{t_B}^t dt'/a(t')$. Here, $t_B$ is a finite time in the past. Now if we designate $t=0$ as the present time, then $t_B$ represents a finite negative value. So the time parameter $t$ covers a range of negative (past) and positive (future) values. Furthermore, defining the comoving hyperspherical coordinate $\chi=\int_0^r\ed r'/\sqrt{1-kr'^2}$, the furthermost radial position from which light is received by a comoving observer at any specific time $t_f$ during the final evolution of universe, is given by
\begin{equation}\label{eq:EventHorizon-general}
\chi_\eh(t)=\int_t^{t_f}\frac{\ed t'}{a(t')}.
\end{equation}  
This is indeed the definition of a cosmological event horizon \cite{Mukhanov2005}. Note that in the case of $t<0$, the value of $t_f$ can be both positive or negative (but of a greater value than $t$). This is because $t_f$ is meant to be located in the future of $t$. \\

Now for a closed universe ($k>0$), $t_f$ would be a positive and fixed value like $t_\textrm{max}$, which is the time during which the universe has experienced its maximum expansion and then has started a re-collapse era. Using Eq.~({\ref{eq:LMP-a(t)}}) the comoving even horizon of a closed \Horava universe becomes
 \begin{equation}\label{eq:EventHorizon-Horava-closed}
\chi_\eh(t)=\sqrt{\frac{3\lambda-1}{k}}\left\{\arctanh\left[\sqrt{\frac{\Lambda_w}{k}}~a(t)\right]-\arctanh\left[\sqrt{\frac{\Lambda_w}{k}}~a(t_\textrm{max})\right]~\right\}.
\end{equation}
This necessitates that for a closed universe it is always $\Lambda_w>0$ and hence, $\lambda>1/3$. The same procedure for an open ($k<0$) \Horava universe (requiring $t_f\rightarrow+\infty$) results in 
 \begin{equation}\label{eq:EventHorizon-Horava-open}
\chi_\eh(t)=\sqrt{\frac{3\lambda-1}{k}}\left\{\frac{\pi}{2}+\arctanh\left[\sqrt{\frac{\Lambda_w}{k}}~a(t)\right]~\right\},
\end{equation}
implying that for an open \Horava universe we have $\Lambda_w<0$ and $\lambda<1/3$. And last but not least, for a flat universe ($k=0$), since we can still consider the case of an infinite expansion within an infinite time in the future, Eq.~(\ref{eq:EventHorizon-general}) yields
 \begin{equation}\label{eq:EventHorizon-Horava-flat}
\chi_\eh(t)=\frac{1}{a(t)}\sqrt{\frac{3\lambda-1}{\Lambda_w}}.
\end{equation}
This includes both cases of $\Lambda_w>0$ with $\lambda>1/3$, and $\Lambda_w<0$ with $\lambda<1/3$. Moreover, defining a physical (areal) hyperspherical coordinate $\bar\chi=a(t)~\chi$ and using it rather than the comoving one, Eq.~(\ref{eq:EventHorizon-Horava-flat}) provides $\bar\chi_\eh=1/H$, which is the de Sitter (or Hubble) horizon.  \\

It is straightforward to show that $\chi_\eh$ constitutes a null hypersurface. For example, for the case of flat universe, we can introduce $\Phi(t,\chi)=\chi-\chi_\eh=0$ and its normal vector
 \begin{equation}\label{eq:normal-EventHorizon}
N_a=\partial_a\Phi(t,\chi)|_{\chi_\eh}=\left(\sqrt{\frac{3\lambda-1}{\Lambda_w}}~\frac{\dot a}{a},1,0,0\right),
\end{equation}
according to which, $N_a N^a=0$.

\subsection{Null Congruence in \Horava Cosmology: The Apparent Horizons}\label{subsec:ApparentHorizon}

In what follows, we take into account a closed universe which in the standard FLRW cosmology is given by $k>0$. So to retain the conformity with the cosmological LMP solutions, we adopt the case of $\Lambda_w>0$ and $\lambda>1/3$. However we let the parameter $\alpha$ change freely from negative values to positive ones. This assumption is crucial to the temporal foliations of the horizon hypersurfaces and the concept of negative time. \\

Having these, Eqs.~(\ref{eq:2-metric}) and (\ref{eq:expansion-general}) yield
\begin{subequations}\label{eq:null-cosmic-expansion} 
\begin{align}
\tu&=\frac{2}{ar}\left(\dot a r+\sqrt{1-kr^2}\right),\label{eq:null-cosmic-expansion-outgoing}
\\
\td&=\frac{2}{ar}\left(\dot a r-\sqrt{1-kr^2}\right).\label{eq:null-cosmic-expansion-ingoing}
\end{align}
\end{subequations}
In a cosmological treatment, we use marginally trapped surfaces for ingoing congruences to foliate cosmological apparent horizons. These are 2-spheres of symmetry characterized by $\td=0$ and $\tu>0$ \cite{Faraoni2013}. In this sense, Eq.~(\ref{eq:null-cosmic-expansion-ingoing}) describes the cosmological apparent horizon on a sphere of the radius
 \begin{equation}\label{eq:cosmological-ApparentHorizon-general}
r_\ah(t)=\frac{1}{a(t)}~\frac{1}{\sqrt{H^2+\frac{k}{a(t)^2}}}.
\end{equation}
It is important to check that unlike event horizons, apparent horizons have not to be null hypersurfaces. In connection with Eq.~(\ref{eq:cosmological-ApparentHorizon-general}), the normal to the apparent horizon hypersurface $\Phi(t,r)=ar-1/\sqrt{H^2+k/a^2}=0$ on the apparent horizon location is
\begin{equation}\label{eq:cosmological-ApparentHorizon-normal-general}
N_a=\left(Hr_\ah\frac{\ddot a}{a^2},~a,0,0\right).
\end{equation}
To talk more strictly about the real physical separation between cosmic objects, it is of benefit to work with the physical (areal) coordinate $\br=a(t)~r$, providing $\br_\ah=1/\sqrt{H^2+k/a^2}$. Together with Eq.~(\ref{eq:cosmological-ApparentHorizon-normal-general}) and after manipulations, we get
\begin{equation}\label{eq:cosmological-ApparentHorizon-normal-norm-general}
N_aN^a=H^2\left[\br_\ah^2-\left(\frac{\ddot a}{a}\right)^2\br_\ah^6\right].
\end{equation}
In a \Horava universe provided by Eq.~(\ref{eq:LMP-cosmological}) and together with Eq.~(\ref{eq:cosmological-ApparentHorizon-general}), we obtain
\begin{equation}\label{eq:cosmological-ApparentHorizon-normal-norm-Horava}
N_aN^a=\frac{4(3\lambda-1)~H^2}{\left(\Lambda_w+3(\lambda-1)\frac{k}{a^2}+\frac{k^2}{\Lambda_w a^4}\right)^3}\left[\left(\frac{\Lambda_w}{2}+\frac{3}{2}(\lambda-1)\frac{k}{a^2}+\frac{k^2}{2\Lambda_w a^4}\right)^2-\left(\frac{\Lambda_w}{2}-\frac{k^2}{2\Lambda_w a^4}\right)^2\right].
\end{equation}
To check the characteristics of \Horava cosmological apparent horizons, we consider points of fixed time $t_0$, on which $\Lambda_w=b\frac{k}{a^2}$ with $b$ to be a positive real number. This way, we can talk about positivity (space-like case) or negativity (time-like case) of the normal vector $N^a$ as a variable depending on both $\Lambda_w$ and the cosmic time $t$. For the special case of $b=1$, we have $N^aN_a=0$ implying a null normal vector and a null apparent horizon. However, this also corresponds to $H=0$, which is relevant to the case of a constant scale factor $a_0=\sqrt{k/\Lambda_w}$ at $t_0\rightarrow-\infty$. On the other hand at negative infinity, Eq.~(\ref{eq:EventHorizon-Horava-closed}) results in $r_\eh\rightarrow+\infty$; the event horizon vanishes. So the event horizon at negative infinity would have been the null apparent horizon $r_\ah(t_0)=1/\sqrt{\Lambda_w}=a_0/\sqrt{k}$. When the apparent horizon is null, it becomes event horizon itself. Note that the negative infinite time is assumed to be the Big Bang time in steady state models.\\

To proceed further, we emphasize that letting $\Lambda_w=bk/a^2$, the term in the brackets in Eq.~(\ref{eq:cosmological-ApparentHorizon-normal-norm-Horava}) has two solutions $\lambda_1=(3-2b)/3$ and $\lambda_2=1-2/(3b)$. Now since $H^2$ and $\br_\ah$ have to be positive, we can categorize $N_a N^a$ in the following ways:

\begin{itemize}

\item{$N_a N^a$ is positive: the apparent horizon is time-like}

\begin{enumerate}
\item{$\Lambda_w>\frac{k}{a^2}$ ~($b>1$)}

\begin{list}{}{}
\item{(a) $\lambda_2<\lambda<1$}

\item{(b) $\lambda>1$}
\end{list}

\item{$\Lambda_w<\frac{k}{a^2}$~~($0<b<1$)}

\begin{list}{}{}
\item{(a) $\lambda_1<\lambda<1$}

\item{(b) $\lambda>1$}
\end{list}

\end{enumerate}


\item{$N_a N^a$ is negative: the apparent horizon is space-like}

\begin{enumerate}
\item{$\Lambda_w>\frac{k}{a^2}$ ~($b>1$)}

\begin{list}{}{}
\item{(a) $\frac{1}{3}<\lambda<\lambda_2$}
\end{list}

\item{$\Lambda_w<\frac{k}{a^2}$~~($0<b<1$)}

\begin{list}{}{}
\item{(a) $\frac{1}{3}<\lambda<\lambda_1$}
\end{list}

\end{enumerate}

\end{itemize}

To clarify the situation, the time-like and space-like apparent horizons have to be plotted as they are evolving in cosmic time. Note that the general case of $\Lambda_w>k/a^2$ can be achieved for every $\alpha>0$ and for all finite negative time and also for $t_0$ within ($0,+\infty$). This means that in a closed \Horava universe, time-like and space-like apparent horizons are available both in the future and in the finite past. Moreover, the case of $\Lambda_w<k/a^2$ corresponds to $\alpha<0$ and also to
$$t_0\leq\frac{\ln\left(\frac{k}{|\alpha|\Lambda_w}\right)}{\sqrt{\frac{\Lambda_w}{3\lambda-1}}}.$$
This means that (if $|\alpha|\Lambda_w>1$) the case of $\Lambda_w<k/a^2$ implies evolution within finite negative time.\\

\begin{figure}
\center{  \includegraphics[width=5cm]{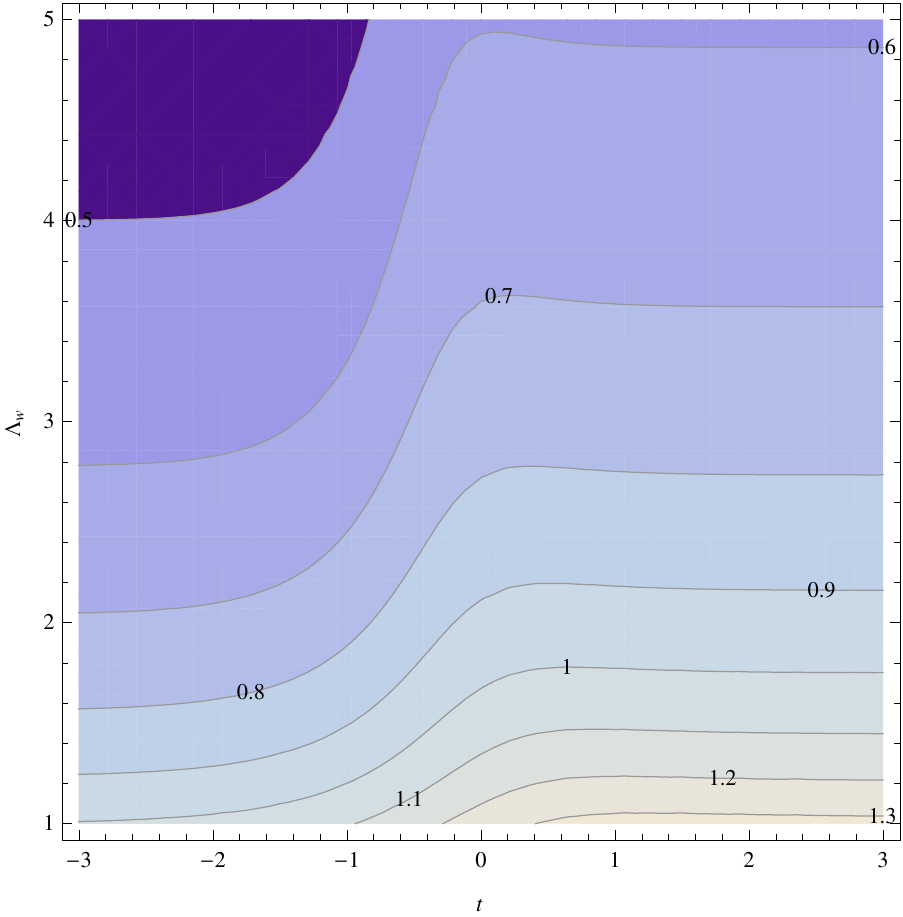}~(a) \hfil
\includegraphics[width=5cm]{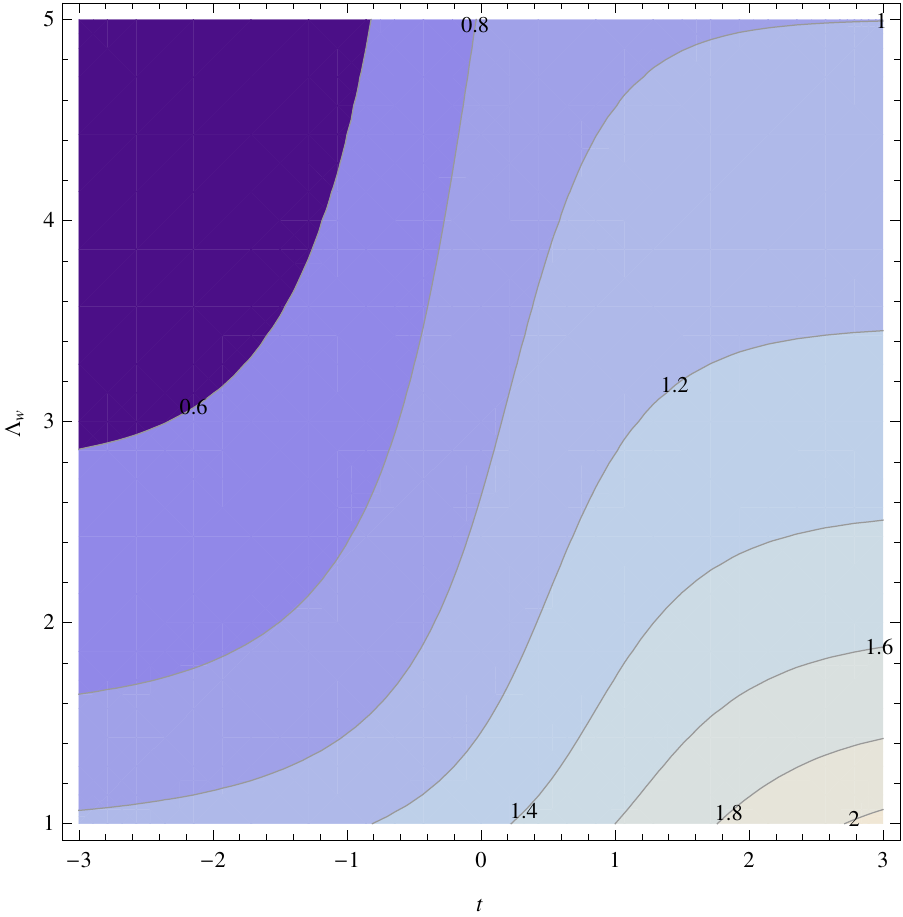}~(b)
\hfil
\includegraphics[width=5cm]{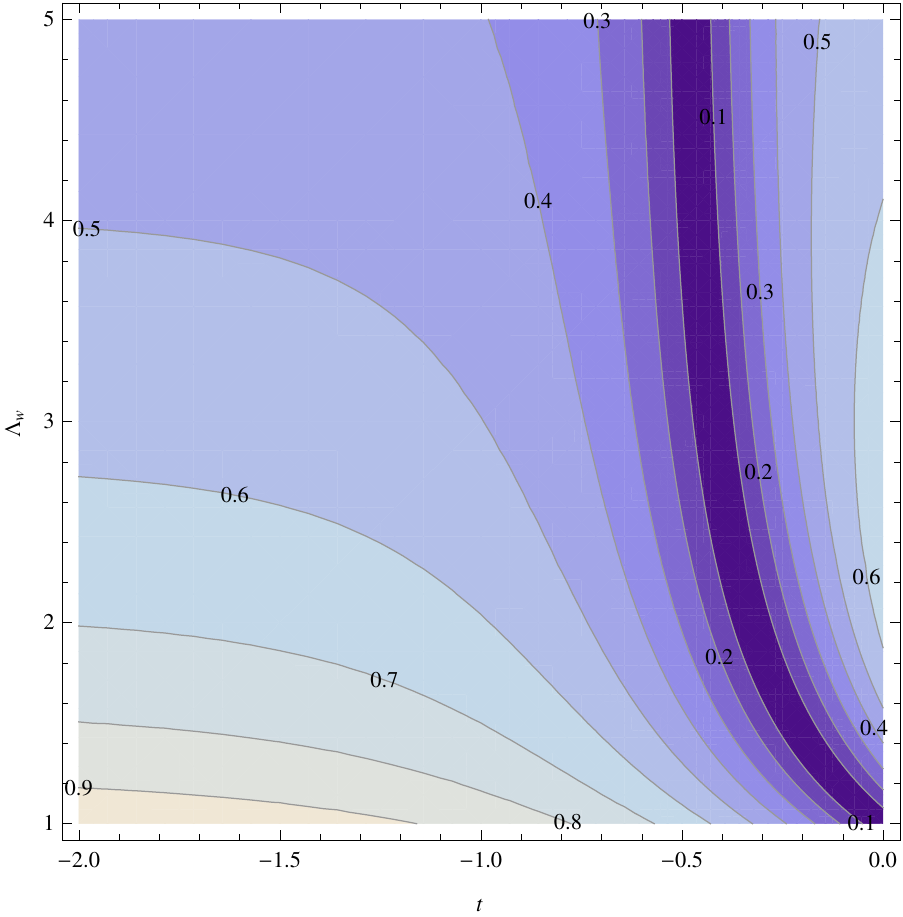}~(c)
\hfil
\includegraphics[width=5cm]{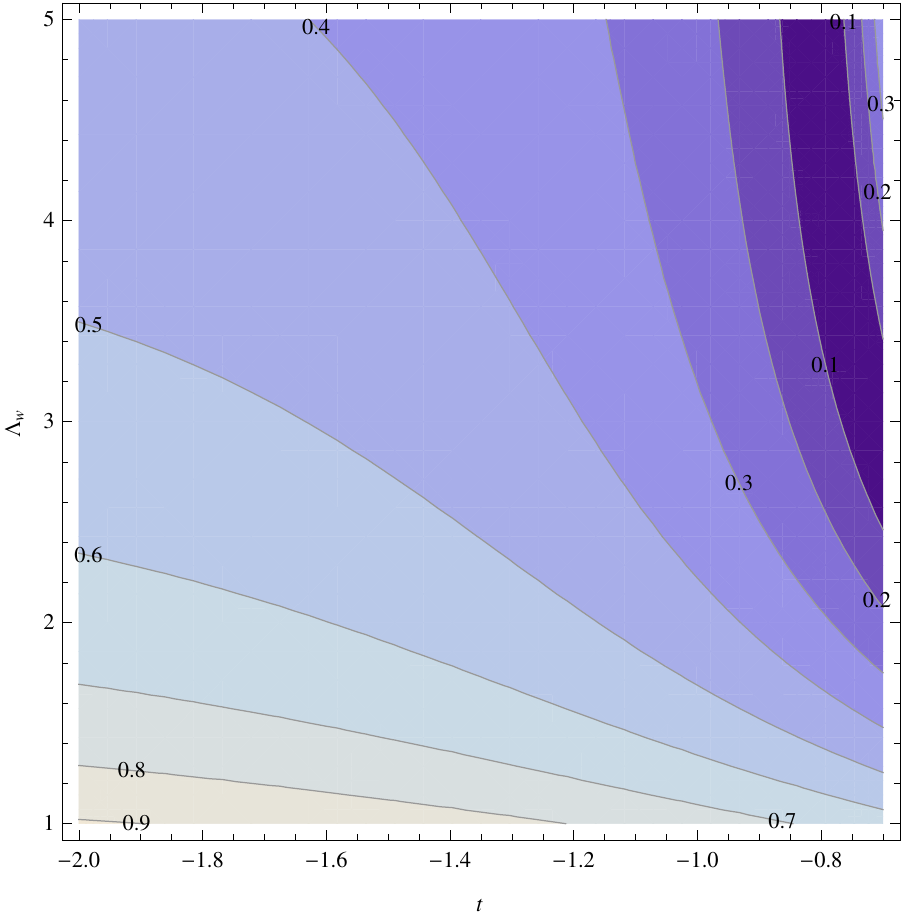}~(d)
\caption{\label{fig:time-like-AH} The evolution of time-like $\br_\ah$ in the future (positive time) and in the past (negative time), in a closed \Horava universe ($k=1$) . At any fixed point $t_0$ within these contours, $\Lambda_w=b\frac{k}{a^2}$, with $b>0$. The dependence on the values of $\Lambda_w$ in these cases corresponds to: (a) $b=2$, $\lambda=\frac{11}{12}$ and $\alpha=1$, (b) $\lambda=2$ and $\alpha=1$, (c) $b=\frac{1}{2}$, $\lambda=\frac{11}{12}$ and $\alpha=-1$, (d) $\lambda=2$ and $\alpha=-1$. For both cases of $\alpha<0$, since $|\alpha|\Lambda_w>1$, the time interval lies within negative values. }}
\end{figure}

\begin{figure}
\center{  \includegraphics[width=5cm]{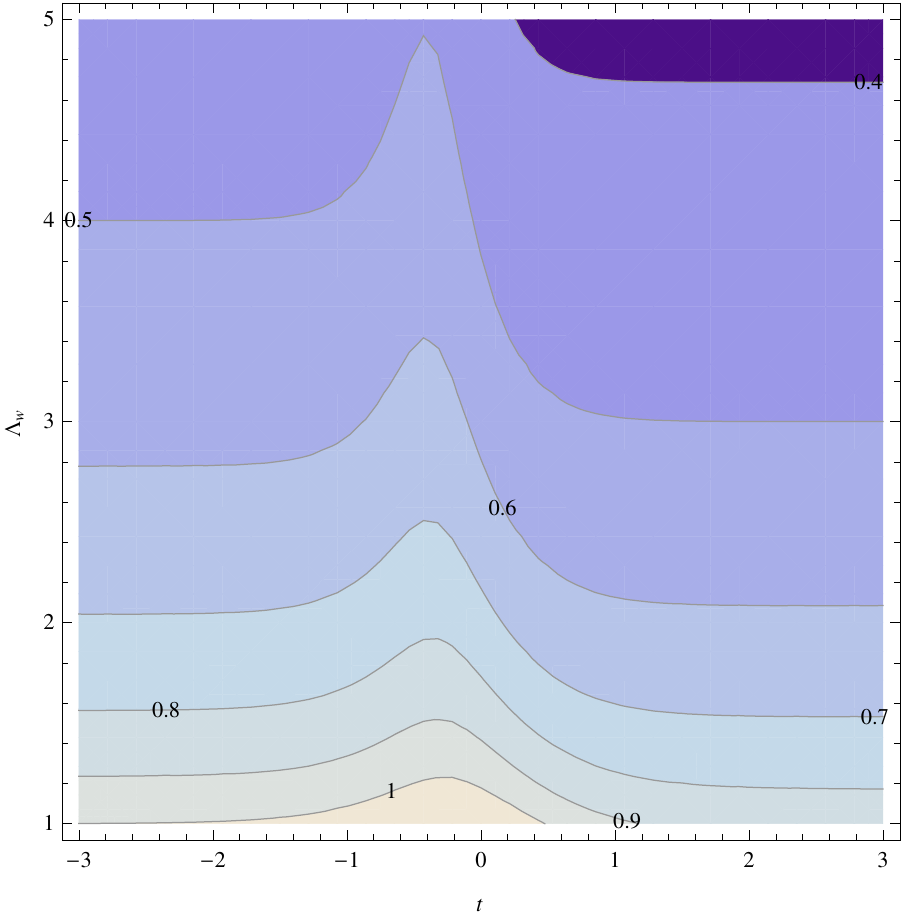}~(a) \hfil
\includegraphics[width=5cm]{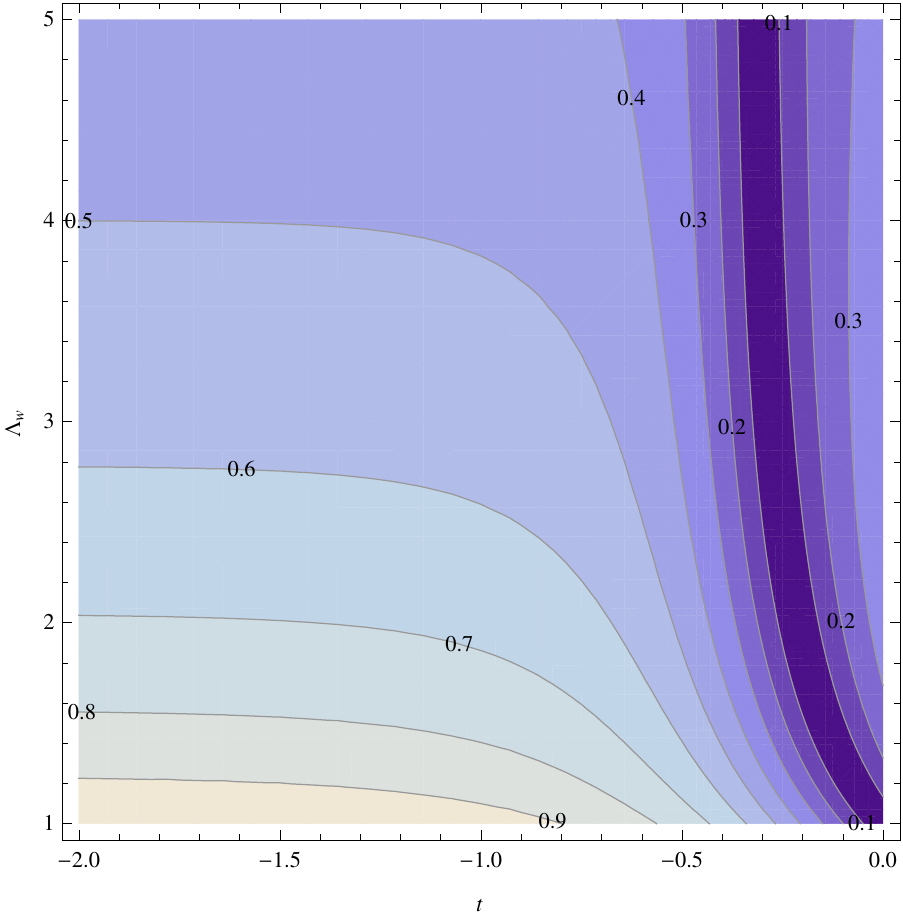}~(b)
\caption{\label{fig:space-like-AH} The evolution of space-like $\br_\ah$ in the future (positive time) and in the past (negative time), in a closed \Horava universe ($k=1$) . At any fixed point $t_0$ within these contours, $\Lambda_w=b\frac{k}{a^2}$, with $b>0$. The dependence on the values of $\Lambda_w$ in these cases corresponds to: (a) $b=2$, $\lambda=\frac{7}{12}$ and $\alpha=1$, (b) $b=\frac{1}{2}$, $\lambda=\frac{7}{12}$ and $\alpha=-1$. For the case of $\alpha<0$, since $|\alpha|\Lambda_w>1$, the time interval lies within negative values. Also according to the case of $\Lambda_w>\frac{k}{a^2}$, we can observe that space-like apparent horizons can form in both the past and in the future. }}
\end{figure}

The evolution of time-like and space-like horizons have been plotted respectively in Fig.~\ref{fig:time-like-AH} and Fig.~\ref{fig:space-like-AH}. Time-like horizons (sometimes called time-like membranes) lie inside the event horizon, therefore  in the interval $\br_\ah<\br<\br_\eh$ we have $\td>0$ and for $\br<\br_\ah$, it is $\td<0$. In contrast, if the apparent horizon is space-like (dynamical horizon), it is $\td<0$ between the two horizons, and it is $\td>0$ beyond the apparent horizon.

\subsection{Trapping Horizons in an Expanding \Horava Universe}\label{subsec:TrappingHorizon-Expanding}

The apparent horizons can be regarded also as trapping horizons if the foliating 2-spheres of symmetry satisfy conditions on the Lie derivative $\lieu\td$ \cite{Faraoni2013}. According to Eqs.~(\ref{eq:Lnull-cosmic-outgoing}) and (\ref{eq:null-cosmic-expansion-ingoing}) and on the apparent horizon $\br_\ah=1/\sqrt{H^2+k/a^2}$, this Lie derivative in a \Horava universe becomes
\begin{equation}\label{eq:Lie-out-Horava}
\lieu\td|_{\br_\ah}=\frac{2}{3\lambda-1}\left(2\Lambda_w+3(\lambda-1)\frac{k}{a^2}\right).
\end{equation}
On the other hand on a cosmological apparent horizon $\td=0$ and $\tu>0$. Since on the apparent horizon Eq.~(\ref{eq:null-cosmic-expansion-outgoing}) reduces to
\begin{equation}\label{eq:null-cosmic-expansion-outgoing-ApparentHorizon}
\tu|_{\br_\ah}=4H,
\end{equation}
this latter condition is automatically satisfied in an expanding universe, where $H>0$. The two quantities in Eqs.~(\ref{eq:Lie-out-Horava}) and (\ref{eq:null-cosmic-expansion-outgoing-ApparentHorizon}) are indeed those which characterize the 2-spheres of symmetry to foliate the trapping horizons. To deal with this, we consider the time-like and space-like apparent horizons and delineate which sort of trapping horizons they are implying. \\

Since in an expanding universe we have always $\tu|_{\br_\ah}>0$, we will encounter past trapped surfaces. Moreover, inside the apparent horizon $\td>0$ and this implies anti-trapped surfaces inside the apparent horizons of an expanding universe; expanding universes behave like white holes, which are regions into which no data can be sent. Outside the apparent horizon, $\td<0$ and therefore there will be un-trapped surfaces. \\

\begin{figure}
\center{  \includegraphics[width=5cm]{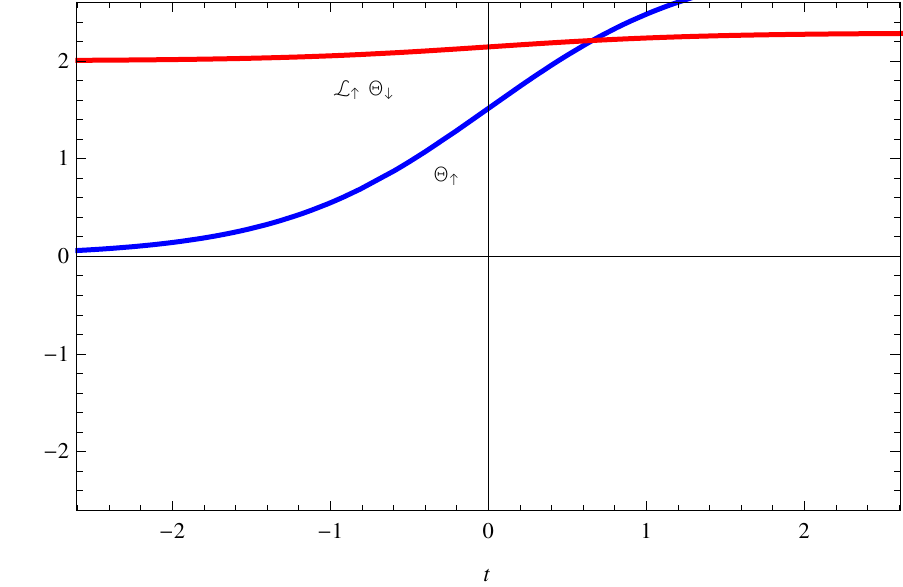}~(a) \hfil
\includegraphics[width=5cm]{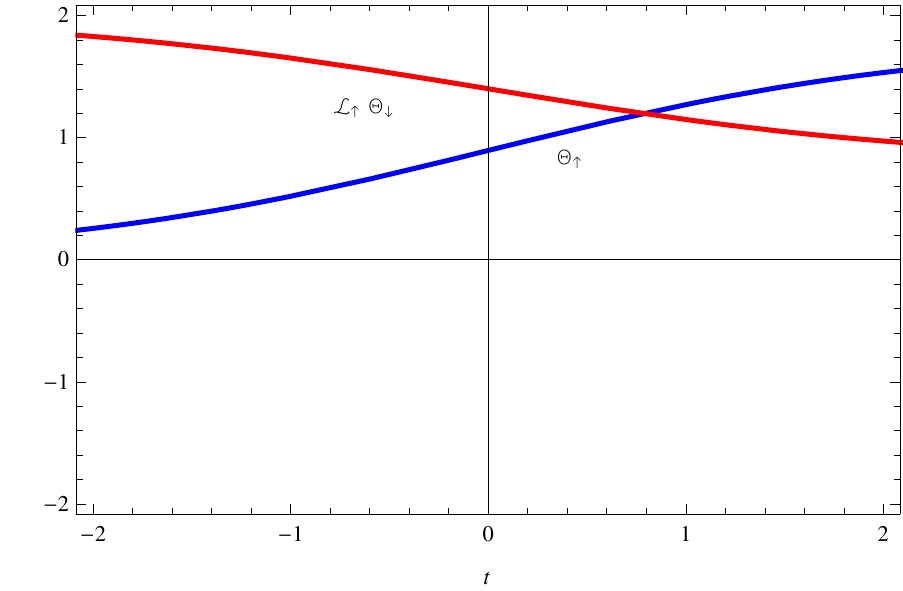}~(b)
\hfil
\includegraphics[width=5cm]{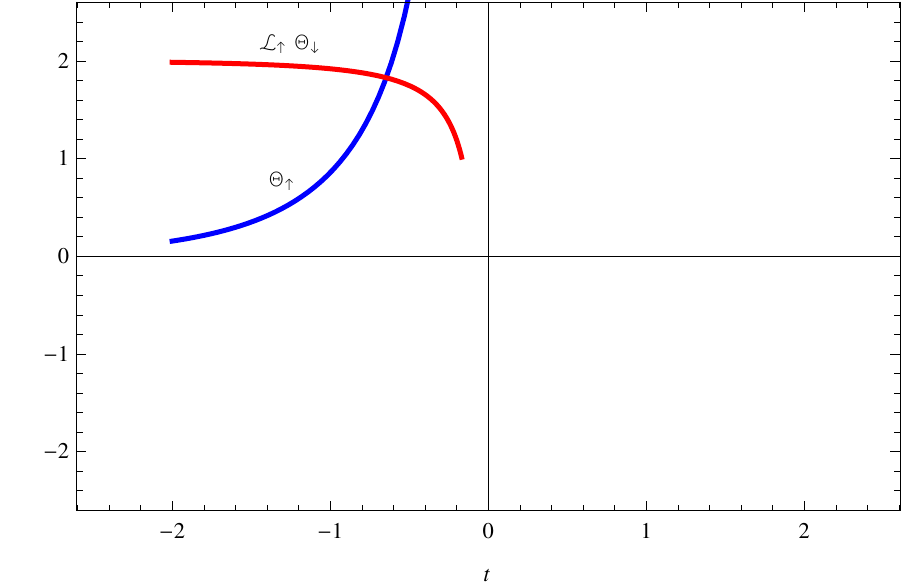}~(c)
\hfil
\includegraphics[width=5cm]{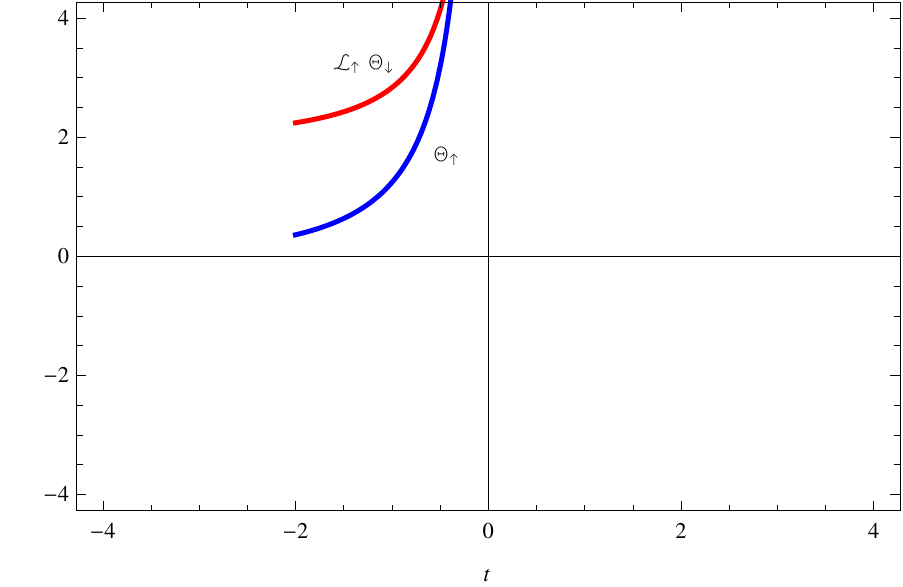}~(d)
\hfil
\includegraphics[width=5cm]{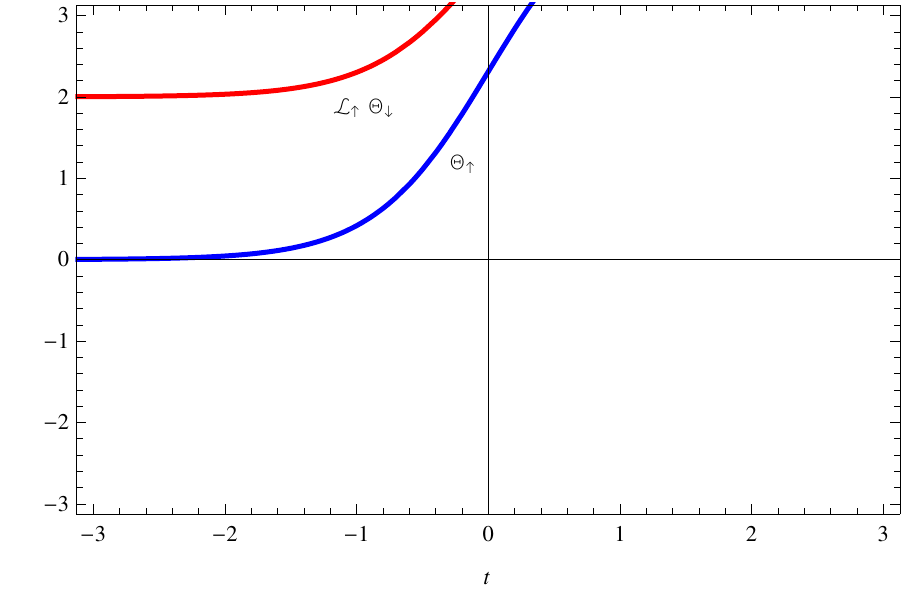}~(e)
\hfil
\includegraphics[width=5cm]{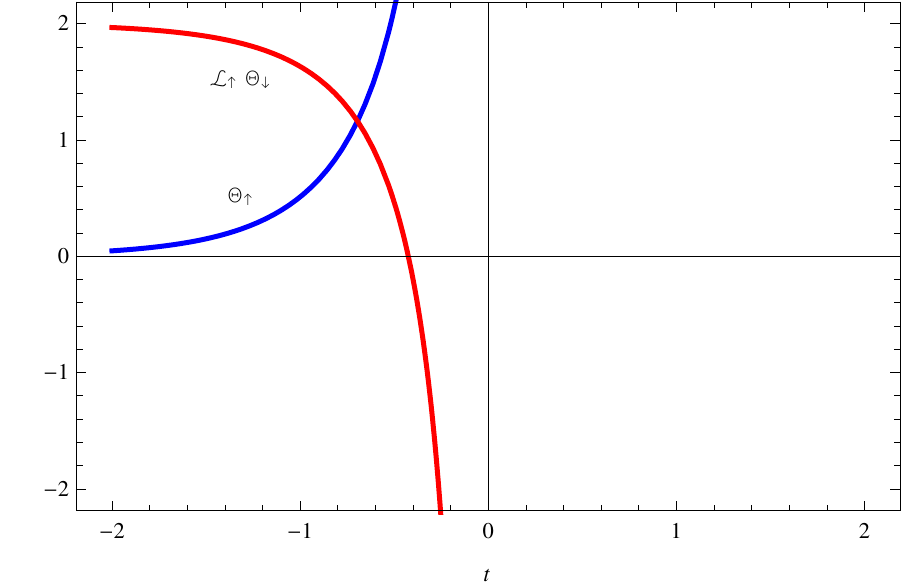}~(f)
\caption{\label{fig:TrappingHorizons-Expanding} The evolution of $\tu$ and $\lieu\td$ on time-like and space-like \Horava apparent horizons, in an expanding closed universe. The diagrams in the cases (a), (b), (c) and (d) correspond exactly to the values of the time-like apparent horizons in Fig.~\ref{fig:time-like-AH}, whereas the cases (e) and (f) are plotted on space-like apparent horizons relevant to those in Fig.~\ref{fig:space-like-AH}. In the first five diagrams $\lieu\td>0$, corresponding to PMITH. However in the last case, we also have some interval where $\lieu\td<0$, which implies PMOTH. }}
\end{figure}

The diagrams in Fig.~\ref{fig:TrappingHorizons-Expanding} demonstrate the evolution of $\tu$ together with $\lieu\td$ on the \Horava time-like  and space-like apparent horizons, in both the finite past and the future. As we can see in the diagrams, for time-like apparent horizons we always have $\lieu\td>0$, therefore those foliating 2-spheres of symmetry are indeed past marginally inner trapped horizons (PMITH) which can form both within positive and negative times (the term ``inner" corresponds to the regions where the observer is located, whereas ``outer" refers to those regions where no observer exists). However there is one occasion in the case of space-like trapping horizons, where $\lieu\td>0$. In such an interval, we will encounter past marginally outer trapped horizons (PMOTH) within negative time. So in this peculiar case, the anti-trapped surfaces can evolve from being inside the apparent horizon, to outside of it. However as we have mentioned before, all cosmological trapping horizons of an expanding universe are indeed horizons in the past, because $\tu>0$ is always true. To avoid any confusion with the positivity and negativity of time, the term ``past" here means that at any fixed time $t_0$ which can be positive or negative, we can observe that outgoing rays started diverging at a time smaller than $t_0$. This makes sense, because the generators $\mlu$ and $\mld$ in Eq.~(\ref{eq:null-cosmic}) are both future directed vectors.

\subsubsection{Contracting Universe Counterpart}\label{subsubsec:TrappingHorizon-Contracting}

The apparent horizons of a contracting universe are given by $\tu=0$ and $\td<0$. Such horizons are also trapping horizons if $\lied\tu\gtrless0$. According to Eq.~(\ref{eq:null-cosmic-expansion-outgoing}) it is obvious that the apparent horizon is still on $\br_\ah=1/\sqrt{H^2+k/a^2}$ and $\td|_{\br_\ah}=4H$, however on the strict understanding that $H<0$. This therefore makes $\td<0$ trivial. On the other hand it turns out that $\lied\tu|_{\br_\ah}$ in a contracting \Horava universe has the same value of $\lieu\td|_{\br_\ah}$ in an expanding one, in Eq.~(\ref{eq:Lie-out-Horava}). Hence, regarding the diagrams in Fig.~\ref{fig:TrappingHorizons-Expanding}, we can infer that all time-like apparent horizons in a contracting \Horava universe are future marginally inner trapped horizons (FMITH), both for positive and negative cosmic time intervals. For the case of space-like apparent horizons it is the same, however in some interval in the case of $\Lambda_w<k/a^2$ and within negative time, we can also encounter future marginally outer trapped horizons (FMOTH) where $\lied\tu<0$. Inside these horizons $\tu<0$, therefore the 2-spheres of symmetry in a contracting universe are trapped surfaces. They are indeed un-trapped outside, where $\td>0$. In this regard, contracting universes behave like black holes.

\section{Summary and Conclusions}\label{sec:conclusion}

The formation of null, time-like and space-like \Horava apparent horizons in black hole and cosmological contexts constitutes the main objective of this paper. Such horizons are generated by sets of ingoing and outgoing congruences. This way we tuned up our method of treatment in configuring 3-dimensional horizons, foliated by 2-dimensional marginally trapped surfaces. For spherically symmetric static \Horava black holes, we investigated KS and LMP solutions. Indeed such astrophysical black holes are causally disconnected from their outer regions by means of 3-dimensional hypersurfaces, on which outgoing congruences cease to expand.  We have shown that KS black holes have two null apparent horizons, however only the outer horizon is also an event horizon. We emphasized the trapped 2-spheres of symmetry inside such an event horizon. Moreover, we indicated that the outer apparent horizon is a FMOTH whereas the inner one is a FMITH. The former corresponds to regions outside which no observer can stand and the latter indicates a horizon inside which there is a time-like region. In the case of LMP static black holes, we encountered only one apparent horizon characterized as a FMOTH.\\

Turning to LMP cosmological solutions, we dealt with a vacuum closed gravitational universe. There we showed that the entire gravitational system is encompassed by distinct event and apparent horizons.  Both types of horizons evolve in time. However, we assigned respective conditions on the Lorentz violating constant and the cosmological term. Such conditions made it possible for the apparent horizons to be both time-like and space-like. According to positivity and negativity of the coupling constant, we inferred that both time-like and space-like horizons could form and evolve in the finite past and in the future. In our approach, the latter is delineated by positive values of the cosmic time, and the former by finite negative time. \\

We also discussed the circumstances for which the \Horava apparent horizons were also trapping horizons. We indicated that the time-like and space-like apparent horizons of an expanding \Horava universe are PMITH, and in some intervals within negative time, PMOTH. Indeed inside these 3-dimensional horizons one would encounter anti-trapped 2-surfaces, implying  white hole regions. Once we turn to a contracting \Horava universe, the trapping horizons become FMITH and within some negative time interval, FMOTH. As a very peculiar case, this latter one is exactly the same trapping horizon as we had around a static spherically symmetric astrophysical black hole. In general, however, contracting universes would behave like black holes since their apparent horizons consist of 3-dimensional hypersurfaces on which the outgoing congruence cease to expand. However, the most important remark is the capability of an empty \Horava universe to generate space-like apparent horizons. In the case of general relativity in an empty FLRW background with a positive cosmological constant $\Lambda$, we get $\br_\ah=1/\sqrt{\Lambda/3}=\textrm{const.}$, implying that the apparent horizon of a vacuum FLRW universe is always null (for all three cases of closed, flat and open universes). However in general relativity, time-like and space-like apparent horizons are only possible for a non-vacuum FLRW universe with $T^{\mu\nu}=\textrm{diag}(\rho, p, p, p)$. Indeed the former is available for $p>-\rho$, and the latter for $p<-\rho$ corresponding to a phantom fluid consisting of an exotic matter. This kind of equation of state for the FLRW universe is capable of creating space-like apparent horizons and also corresponds to a super-acceleration. On the other hand, we found that even an empty \Horava universe can create space-like apparent horizons, if it has been  configured properly. Accordingly, an empty \Horava universe with $\lambda>\lambda_1$ in the past and/or $\lambda>\lambda_2$ in both the past and in the future, corresponds to a FLRW universe (in general relativity) with $p>-\rho$ with non-exotic matter. Similarly, the same \Horava universe with $1/3<\lambda<\lambda_1$ and/or $1/3<\lambda<\lambda_2$ corresponds to FLRW with exotic matter. It is worth recalling that null apparent horizons in an empty \Horava universe were possible only in the infinite past. \\

Furthermore, since each specifically configured \Horava universe can envisage only one kind of apparent horizon, one can infer that these horizons cannot occur together. However as we discussed, they can all evolve in time. This may be regarded as a significant feature of the \Horava cosmological black/white holes which we have discussed in this paper.

\end{document}